\input psfig.sty

\documentclass{aa}
\begin{document}

   \thesaurus{08.16.7 Crab, 05.01.1}  
   
   \title{A New HST Measurement of the Crab Pulsar Proper Motion.
   \thanks{Based on observations with the NASA/ESA
Hubble Space Telescope,
obtained at the Space Telescope Science Institute, which is operated
by AURA, Inc., under NASA contract NAS 5-26555}}

\author{P. A. Caraveo\inst{1,2} and R.P. Mignani\inst{3}} 

   \offprints{P.A. Caraveo}

   \institute{Istituto di Fisica Cosmica del CNR, Via Bassini 15, 20133 
   Milan, Italy - 
            email: pat@ifctr.mi.cnr.it 
         \and 
	     Istituto Astronomico, Via Lancisi 29, 00161 Rome, Italy 
	\and
	STECF-ESO, Karl Schwarzschild Str.2  D85740 Garching, Germany -
         email: rmignani@eso.org
         }

 \date{Received ; accepted  }

\titlerunning{Crab Proper Motion}
\maketitle

\begin{abstract} We have used a set of archived HST/WFPC2 exposures
of the inner regions of the Crab Nebula to obtain a new measurement of
the pulsar proper motion, the first after the original work of Wyckoff
\&  Murray, more  than 20  years ago.  Our  measurement   of the pulsar
displacement, $\mu   = 18 \pm 3$  ~mas~yr$^{-1}$,  agrees well  with the
value obtained previously.  This is noteworthy, since the data we have
used span less than 2 years, as opposed to the  77 years required in the
previous work. \\
With a position angle of $292^{\circ} 
\pm 10^{\circ}$, the proper motion vector 
appears 
aligned with the axis of
symmetry of the inner Crab nebula, as defined by  the direction of the
X-ray jet observed by ROSAT.  Indeed, if the  neutron star rotation is
to be held responsible  both for the X-ray jet   and for the  observed
symmetry, the  Crab
could provide
an  example of alignment between
spin axis and proper motion. 
   
   \end{abstract}

%

\section{Introduction}
The Crab pulsar has  been the second  pulsar to be associated with its
supernova remnant (Comella  et al. 1969)  and  the interaction between
the two has been subject of deep and detailed studies (e.g. Kennel \&
Coronoti  1984).   Recently, associating the
ROSAT/HRI picture of the pulsar and its surroundings with HST/WFPC2 images
of the remnant,  Hester et  al.  (1995)  have   drawn a
convincing picture of the central part of  the Crab Nebula "symmetrical
about the  (presumed) rotation axis of the  pulsar" 
with such an axis lying  "at an 
approximate position angle of $115 ^{\circ}$ east of north". 
However, linking the  pulsar rotation to the
remarkably symmetrical appearance of the high resolution X and optical
data  is clearly a  kind of default solution  since "the only physical
axis that exists for the pulsar is its spin axis". Although generally
correct, this statement may not represent a complete description 
of the Crab pulsar, which is known to move.
Here we want to draw attention to the possible relationship between the
Crab pulsar proper motion and the symmetric appearence of the inner Crab 
Nebula. 
\\
Isolated Neutron Stars
are fast moving objects (e.g. Caraveo 1993; 
Lyne \& Lorimer 1993; Lorimer 1998), and
the Crab is no exception.  Measurements of the proper motion of Baade's
star (later recognized  to  be the   optical counterpart of  the  Crab
pulsar)  were  attempted several  times  (e.g. Trimble 1968), yielding
vastly different  values. This prompted  Minkowski  (1970) to conclude
that the proper  motion of the  star was not  reliably measurable.  
\\
The situation  changed few  years  later, when  Wyckoff \& Murray  (1977)
obtained  a new value   of the  Crab proper   motion which  allowed to
reconcile  the pulsar  birthplace   with  the  center of   the nebula,
i.e. the filaments' divergent point. The {\it relative} proper motion, 
measured 
by
Wyckoff \& Murray over a time span of 77 years, amounts to a total yearly 
displacement of 
$15 \pm 3$ ~mas, corresponding to a transverse velocity of 123 km~s$^{-1}$ for
a pulsar distance of
2 kpc. 
\\
What matters here is the direction of such a motion, i.e. its
position angle of $ 298^{\circ} \pm 10^{\circ}$.
Taken at face value, this direction  is certainly compatible 
with the Crab axis of symmetry, defined by Hester et al. (1995),
hinting an alignement between the pulsar proper motion and the major
axis of the nebula.
\\
Given   the non  trivial
consequences  of  this  evidence,  we  have    sought an
independent  confirmation of the pulsar  proper  motion. Owing to the
dramatic evolution of telescopes as well as optical detectors in the last 20
years, we are now in a position to measure anew the Crab proper motion
in a time span  much shorter than the 77  years required by  Wyckoff \&
Murray.


\section{Defining the Data Set}
Proper motion measurements rely on  accurate relative astrometry.   In
order to measure the tiny angular displacement  of the Crab pulsar, we
need high resolution images taken at  different epochs. Currently, the
best instrument to pursue this task is the Wide Field Planetary Camera
2 (WFPC2), onboard the Hubble Space Telescope.  Luckily enough, the Crab pulsar
is a conspicuous target so that, since the first telescope refurbishing, it
has been repeatedly observed (Hester et  al. 1995, 1996;
Blair et al.   1997).  Of course,  different observers  used different
filters and placed the pulsar either in one of the Wide Field Camera 
(WFC) chips or, more often, in the Planetary Camera (PC). 
\\
In order to define a
homogeneous data  set,  first we have gone through the  exposure list  to
single out images obtained through the same filter.  Since
the 547M medium bandpass ($\lambda= 5454\AA; 
\Delta \lambda=486.6\AA$) turned out to  be the most frequently  used,
we have examined all the images taken through this filter.  The {\rm 547M} data set  (listed  in Table 1) has   been retrieved from the   HST
public database, and, after combining and cosmic ray cleaning, 
all images have been
inspected  to  define  a  suitable set  of   "good  quality" reference
stars.

\begin{table}
\begin{tabular}{|l|l|l|l|} \hline

{\em Obs.} & {\em Date}           & {\em Camera} & {\em Exp. (s)  } \\ \hline

1  & Mar $9^{th}$  1994  & WFC\#2  & 2\,000   \\

2  & Jan $7^{th}$  1995  & WFC\#3  & 3\,600   \\

3  & Aug $15^{th}$ 1995  & PC      & 2\,000   \\

4  & Nov $6^{th}$  1995  & PC      & 2\,000   \\

5  & Dec $29^{th}$ 1995  & PC      & 2\,000   \\

6  & Jan $20^{th}$ 1996  & PC      & 2\,000   \\ 

7  & Jan $26^{th}$ 1996  & PC      & 2\,000   \\

8a & Jan $26^{th}$ 1996  & WFC \#2 & 2\,000   \\ 

8b & ...                 & WFC \#3 & 2\,000   \\ 

8c & ... 		 & WFC \#4 & 2\,000   \\ 

9  & Feb $2^{nd}$ 1996   & PC      & 2\,000   \\ 

10 & Apr $16^{th}$ 1996  & PC      & 2\,000   \\ \hline
\end{tabular}

\label{} 
\caption{547M filter data set selected from the Crab observations
available in the HST public database. For each exposure we list the
date, the chip containing the pulsar and the exposure time in seconds.}
\end{table}
 
When doing astrometric studies, the presence of good reference
stars is very  important.  An outstanding  image  without  at least  4
reference  objects, chosen to be well  below the  saturation limit, but
bright enough to yield sufficient counts for precise positioning, 
is of no use for our purposes. This is particularly true for the Planetary 
Camera which, in spite of its   
much sharper angular
resolution (0.0455 arcsec/px as opposed to 0.1 arcsec/px of the WFC),
suffers from the limited dimensions  
(35  $\times$  35 arcsec)   of its field of view.
Indeed,  among the several PC observations  listed in Table 1,
only \#3, which is shown in Fig.1, contains 4 reference stars. \\
Thus,  only observations \#1,2,3 and  8, covering a time span of
1.9 years, appear suitable for our  astrometric analysis. 

\section{The Relative Astrometry} 

Precise alignement of these images is our next task.
The traditional astrometric approach would call for a linear 
image-to-image trasformation,  requiring at least four constants, namely 
two independent ($x$ and $y$) translations, rotation and image scale.
However, since the paucity of reference stars
would have hampered the overall accuracy of the superposition, 
we applied the {\it rotate-shift} procedure devised by Caraveo et al. (1996)
in order to reduce the plate constants to be computed.
This method takes advantage of the accurate mapping of the geometrical 
distorsion of the WFPC2 to define the instrument scale, while
the telescope roll angle is used to {\it a priori} align our images in RA and 
DEC.
Thus,  the statistical weigth of the few common stars is used only to
compute the $x$ and $y$ shifts.\\
Therefore, our alignement recipe is as follow:\\
${\bullet}$ -- first, the frames have  been  corrected for  the WFPC2
geometrical distorsion (Holtzman et  al. 1995) using the {\it wmosaic}
task    in {\it STSDAS},  which  also  automatically  applies the scale
transformation  between the PC  and WFC  chips.  As a result, all  the
"corrected"   images have the  same pixel  size,  corresponding to 0.1
arcsec (i.e. 1 WFC  px); \\ 
${\bullet}$ -- second, the  frames have been aligned in right
ascension  and   declination according  to their roll angles;\\
${\bullet}$ -- third, the
"best"  positions  of the Crab   pulsar, as well  as those of  the reference
stars, have been  computed by 2-D gaussian  fitting of their profiles,
using specific  MIDAS  routines.   Particular care  was  used  for the
pulsar itself, in order to make sure  that the object's centroid is not
affected by the emission knot observed $\sim$ 0.7 arcsec to the SE.  A
positional accuracy  ranging from 0.02 px to  0.03 px was achieved for
the pulsar ($V = 16.5$) as well as for the reference stars ($17 \le V \le
19$).
It is worth noting that this result is by no means an exceptional one;
Ibata \& Lewis (1998) obtain similar accuracies for significantly 
fainter objects.\\ 
${\bullet}$ -- Finally, we used the common reference
stars  (1 to 4 in   Fig. 1) to compute  the  linear shifts needed  to
overlay the  different frames  onto image  \# 1, which   was used as a
reference.  This  procedure did not always  achieve the same degree of
accuracy.   While obs.\#3-to-obs.\#1   and    obs.\#8a-to-obs.\#1  yielded
residuals close to 0.04  WFC pixels, the superpositions involving
obs.\# 2 and \#8b,c resulted in higher residuals ($\ge$ 0.1~px).
Unfortunately, we
cannot offer an explanation for this effect, other than noting that it
arises when comparing images obtained with different chips of the Wide
Field Camera. Therefore, we were forced to reduce our data set to just
one  Wide Field chip. Since we  had no a priori  reason  to prefer one
particular chip,  we selected the chip which  maximized the time span.
This turns out to be chip\#2, with obs.\#1 and \#8a. To these, 
PC observation  \#3 can  be added. 
These three images, accurately superimposed, are our final data set.

\begin{figure}
\centerline{\hbox{\psfig{figure=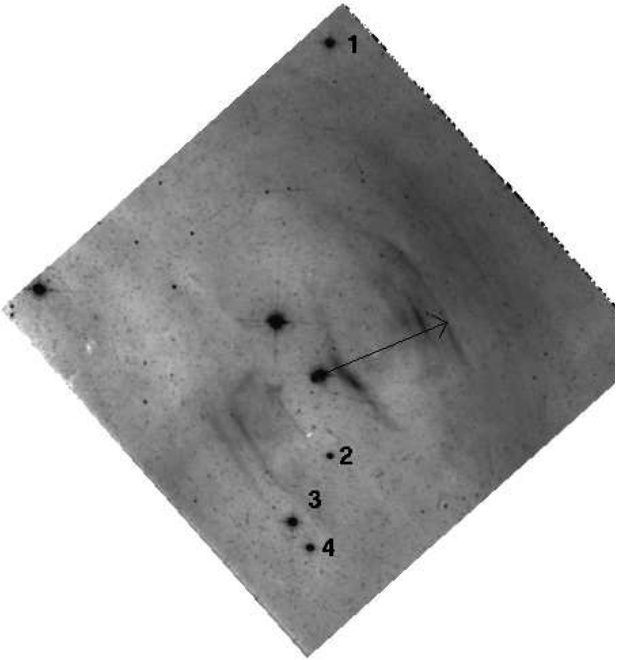,height=9cm,clip=}}}
\caption{2\,000 s PC image of the inner Crab Nebula 
taken  in August  1995 through the   F547M filter (obs.\#3 in  Table 1).
North to the top, East  to the left.  The  labels mark the stars  used
for relative astrometry. The arrow shows the Crab pulsar proper motion
direction.} 
\end{figure}

\begin{figure}
\centerline{\hbox{\psfig{figure=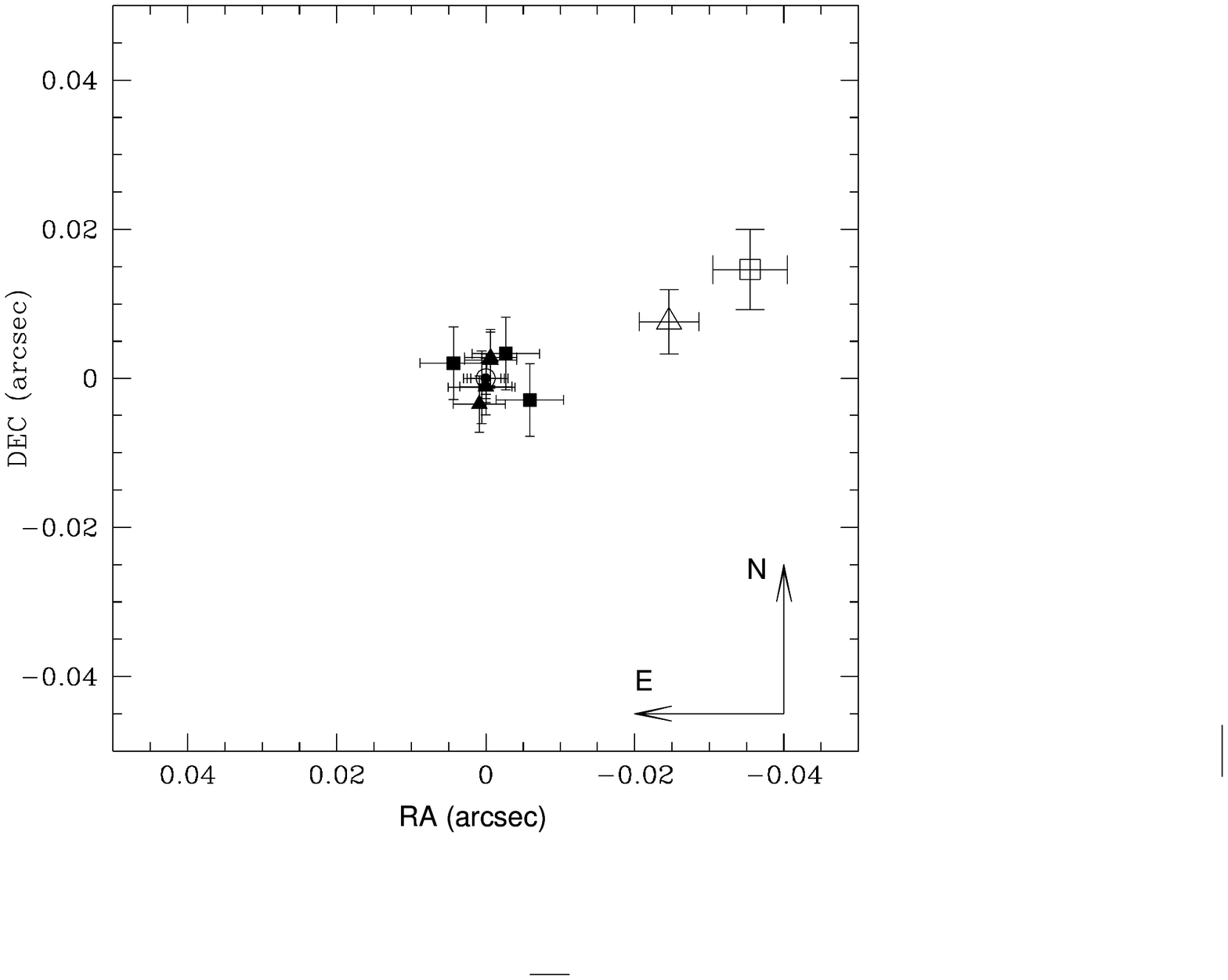,height=8cm,clip=}}}
\caption{Relative positions measured for the 4 reference stars (filled symbols)
as well as for the Crab pulsar (open symbols) in the 
$(\alpha,\delta)$ plane measured at three different epochs
(circle:1994, always in the 0,0 point, since the 1994 observation
has been used as our reference; triangle:1995; square:1996). Axes in arcsec.}
\end{figure}

\section{Results}
It  is now possible  to compare  the  positions obtained  for the Crab
pulsar over 1.9 years. This is done in Table 2, where we give positions,
relative displacements and errors,
measured for the Crab pulsar as well as for the four 
reference stars.
While 
the positions measured for the Crab pulsar in obs.\#3 and \#8a 
show a small variation,
marginal in $y$ but certainly significant in $x$, no significant
displacement is seen in any of the reference objects. 
This is
shown in Fig. 2, where we  have plotted in the $(\alpha,\delta)$ plane
the coordinate offsets measured  in obs.\#3 and  \#8a wrt  obs.\#1,
which represents the 0,0 point in the  figure.  While the positions of
the  reference  stars at   the  three different   epochs are virtually
unchanged, the  pulsar is clearly  affected by a proper motion
to  NW.  A  linear  fit to  the ${\alpha}$ and  ${\delta}$
displacements yields the Crab proper motion  relative to the reference
stars. This turns out to be

\begin{center}
$\mu_{\alpha}=-17   \pm  3$     ~mas~yr$^{-1}$,   $\mu_{\delta}=  7   \pm
3$ ~mas~yr$^{-1}$
\end{center}

\noindent
corresponding  to  an overall  annual  displacement  $\mu  = 18  \pm 3$
~mas~yr$^{-1}$  in  the plane  of  the sky,  with  a position  angle of
$292^{\circ} \pm 10^{\circ}$.  This vector  is also shown in
Fig.1. \\ 
Our result is to be compared with the value of 
  
\begin{center}
$\mu_{\alpha}=-13 \pm 2$~mas~yr$^{-1}$, $\mu_{\delta}= 7 \pm 3$~mas~yr$^{-1}$
\end{center}

\noindent
with   a position angle  of  $298^{\circ} \pm  10^{\circ}$, obtained by
Wyckoff and Murray over a time span of $\simeq$ 77 years. 
 
\begin{table*} 
\begin{center}
\begin{tabular}{|l|l|l|l|l|l|} \hline
{\em Obj} & {\em Obs.\#} & {\em x} & {\em $\Delta x$} & {\em y} & {\em $\Delta 
y$} \\ \hline 
Crab & \#1  & -167.42 (.03) & - 		 &  92.84  (.03) & - \\ 
     & \#3  & -167.19 (.04) & +0.23 (.05) &  92.91  (.04) & +0.07 (.05) \\
     & \#8a & -167.09 (.05) & +0.33 (.07) &  92.98  (.05) & +0.14 (.07) \\ 
\hline
1    & \#1  & -160.80 (.02) & - 		 & 334.59  (.02) & - \\
     & \#3  & -160.80 (.04) &  0.00 (.05) & 334.58  (.04) & -0.01 (.05)\\
     & \#8a & -160.80 (.05) &  0.00 (.07) & 334.58  (.05) & -0.01 (.07) \\ 
\hline
2    & \#1  & -160.70 (.03) & - 	 	 &  33.59  (.03) & - \\ 
     & \#3  & -160.69 (.04) & +0.01 (.05) &  33.62  (.04) & +0.03 (.05)\\
     & \#8a & -160.67 (.05) & +0.03 (.07) &  33.62  (.05) & +0.03 (.07) \\ 
\hline
3    & \#1  & -188.43 (.02) & - 		 & -14.64  (.02) & - \\
     & \#3  & -188.42 (.04) & +0.01 (.05) & -14.62  (.04) & +0.02 (.05)\\
     & \#8a & -188.37 (.05) & +0.06 (.07) & -14.67  (.05) & -0.03 (.07) \\ 
\hline
4    & \#1  & -175.42 (.02) & - 		 & -33.30  (.02) & - \\
     & \#3  & -175.43 (.04) & -0.01 (.05) & -33.32  (.04) & -0.02 (.05)\\
     & \#8a & -175.46 (.05) & -0.04 (.07) & -33.28  (.05) & +0.02 (.07)  \\ 
\hline
\end{tabular}

\end{center}

\caption{
Pixel coordinates $(x,y)$ and relative displacements $(\Delta x,\Delta
y)$ measured at three different epochs for the Crab pulsars as well as
for the four  reference stars.  All frames have  been aligned in right
ascension and  declination   ($x$ and   $y$ increasing  Westward   and
Northward)  before being overlaid  to obs.\# 1.   While for obs.\# 1
the  errors   are due  only to   the  centroid fitting  algorithm, for
obs.\#3  and \#8a they also  include the uncertaity  arising from the
image superposition (0.03 and 0.04 px for obs.\# 3 and \# 8a,
respectively).} 
\end{table*}

\section{Conclusions}
With  two   independent, fully consistent   measurements,   we  can  now
proceed to
compare the Crab pulsar proper motion direction with the axis of
symmetry of the inner nebula.  
This task is an easy one, since we can use figure 8 of Hester
et al. (1995), where such axis is coincident with the direction defined by
the Knot1-Knot2 alignment.  According to Hester et al. the position
angle of
this direction  is   $\simeq  115^{\circ}$, to  which   an   offset  of
$180^{\circ}$ is to be added to take into account the direction of the
Crab motion. This yields a value of $\simeq  295^{\circ}$, to be compared to
our value  $\simeq 292^{\circ}$ or to  that of Wyckoff \& Murray ($\simeq
298^{\circ}$). 
\\
Although all these values are affected by non negligible errors,
both known, as in the case of the proper motion, and unknown, as in the 
case of the roughly defined axis of symmetry,
an alignement between the pulsar proper motion and the "axis of symmetry" of 
the inner nebula seems to be present.
In fact, 
the experimental evidence gathered so far 
shows that the Crab pulsar is moving along the major axis of the Crab Nebula
(Wyckoff \& Murray 1977)
and that both  
the knots and the X-ray jet 
appear 
aligned to the pulsar
proper motion. 
Although
the significant uncertainties of the relevant parameters
leave open the possibility of a chance coincidence, 
it is interesting to speculate on the 
implications of such an alignement. \\
Since a neutron star acquires its proper motion at birth,
there is no doubt that the pulsar motion has been present "ab initio",
well before both knots and jets came into existence. However, the proper motion
energy content is
far too small to account for the surrounding structures and their rapid
evolution. Therefore, the link, if any, between proper motion and axis of
symmetry must be through some basic characteristics which was
also present when the Crab pulsar was born. 
Hester et al. (1995) proposed a scenario associating the symmetrical
appearence of the Nebula with the pulsar spin axis.
Under this hypothesis, the neutron star motion would turn out to be aligned 
with the spin axis, reflecting an asymmetry of the supernova explosion
along the progenitor's spin axis.
Proper motion spin axis alignements have been discussed in the literature
(see e.g. Tademaru 1977), but no conclusive evidence was found.
\\
If the X-ray jets do indeed trace the pulsar spin axis, 
and the relation between proper motion and axis of symmetry is not
a fortuitous one, 
the Crab would provide
the first example of such an alignement. While this would shed some
light on the mechanisms responsible for the pulsar kick (e.g. Spruit \& Finney
1998), one must immediately add that nothing similar has yet been found for 
the very
limited sample of the young pulsars we know.
PSR 0540$-$69, the twin of the Crab in the Magellanic Cloud, is too far
to allow for proper motion measurements in any reasonable time span. 
PSR 1509$-$58 does not have
a definite optical counterpart. The significantly older Vela
pulsar does not show any alignement between its 50 mas/y proper motion
(Nasuti et al. 1997) and the X-ray jet proposed by Markwardt \& \"Ogelman 
(1995).
\\
Before speculating any further,
better data are needed to improve our knowledge on the geometry of the
Crab pulsar  surroundings.One more  HST observation could  easily
improve the determination of the proper motion position angle.  A very
accurate  proper motion   measurement, however, will   not  settle the
problem without a substantial improvement on the X-ray side. A sharper
X-ray image  is  needed to better assess   the position angle  of  the
jet(s) together with their shape and overall dimension.  The AXAF High
Resolution Camera could improve significantly on  the fuzzy picture of
the inner Crab Nebula obtained by ROSAT. \\
Irrispective  of  future  developments, however, the  presence  of  an observed
motion  adds a definite   direction  to the cilindrically  symmetrical
appearence of the Crab. 

      

\end{document}